\documentclass[11pt]{article}
\usepackage{indentfirst}
\usepackage{cite}
\usepackage{rotating}
\usepackage{amsfonts}
\usepackage{amsmath}
\usepackage{amssymb}
\usepackage{graphicx}
\usepackage[colorlinks=true,linkcolor=red,citecolor=blue]{hyperref}
\usepackage{url}
\usepackage{color}
\usepackage{slashed}
\usepackage{multirow,booktabs}
\begin{document}
\title{Global Analysis of Charmless $B$ Decays into Two Vector Mesons in Soft-Collinear Effective Theory}
\author{ Chao Wang$^{1}$,~~ Si-Hong Zhou$^{1}$,~~ Ying Li$^{2}$\footnote{Email:liying@ytu.edu.cn},~~
Cai-Dian L\"u$^{1}$\footnote{Email:lucd@ihep.ac.cn}\\
{\small \it 1.~Institute of High Energy Physics, CAS, P.O. Box 918, Beijing 100049, China  }\\
{\small \it and School of Physics, University of Chinese Academy of Sciences, Beijing 100049, China} \\
{\small \it 2.~Department of Physics, Yantai University, Yantai 264005, China}\\
}
\maketitle
\begin{abstract}
Under the framework of  soft-collinear effective theory, we analyze  the charmless $B\to VV$ decays in a global way at leading power in $1/m_b$ and leading order in $\alpha_s$ with $V$ denoting a light vector meson.  In the flavor SU(3) symmetry, decay amplitudes for the 28 decay modes are expressed in terms of  8 nonperturbative parameters.  With 35 experimental results, we fit these 8 nonperturbative parameters.  Annihilation contributions are neglected due to power suppression in the $m_b\to \infty$ limit, so we include in the fit the nonperturbative   charm penguins, which  will play an important role  in understanding  the direct {\it CP} asymmetries. Charming penguins are also responsible for the large transverse polarizations of penguin-dominated  and color-suppressed decays. With the best fitted  parameters, we calculate  all possible physical observables of 28 decay modes, including   branching fractions,   direct {\it CP} asymmetries, and the complete set of polarization observables. Most of our results are   compatible with the present experimental data when available, while the others  can  be examined  on  the ongoing LHCb experiment and the forthcoming Belle-II experiment. Moreover, the agreements and differences with results in QCD factorization and perturbative QCD approach are also discussed. A few    observables are suggested to discriminate  these   different approaches.
\end{abstract}
\newpage

\section{Introduction}\label{sec-1}
Weak decays of  $B$ mesons can  provide us not only an interesting avenue to understand the $CP$ violation and the quark flavor mixing in the standard model (SM), but also a powerful means to probe    new physics effects beyond the SM \cite{Buchalla:2008jp, Antonelli:2009ws}. Experimentally,  the $B$ meson weak decays have been extensively investigated in the past decades  at both $e^+e^-$ colliders ($B$-factories) at SLAC and KEK, and  hadronic environments such as Tevatron. The experimental  results obtained by   Babar, Belle, D$0$ and CDF collaborations have reached an unprecedented precision \cite{Bevan:2014iga}. In recent years, with the running of LHC at CERN, the LHCb experiment has become the main player of the bottom quark. To date a large amount of  bottom meson  and baryon events has been accumulated, and a number of results related to non-leptonic decays have been released and more  analyses are ongoing \cite{Bediaga:2012py}. In addition, the forthcoming Belle-II experiment  will  further improve the experimental precision  significantly \cite{Aushev:2010bq}.

With the plethora of precise experimental data on nonleptonic decays at hand, theoretical predictions at the same level of accuracy are very much desired. However, for dealing with the hadronic $B$ decays theoretically, the major obstacle is   to reliably evaluate the hadronic matrix elements of local operators between the initial and final hadronic states, especially due to the nontrivial QCD dynamics involved. In the past two decades, three major QCD-inspired approaches had been proposed to calculate the charmless non-leptonic $B$ decays, namely,   QCD factorization (QCDF) \cite{Beneke:2000ry}, perturbative QCD (PQCD) \cite{Lu:2000em}, and soft-collinear effective theory (SCET) \cite{Bauer:2000yr}. A prominent  difference among them resides in the treatment of dynamical degrees of freedom at different mass scales, namely the power counting. Despite of this difference, all of them are based on power expansions in $\Lambda_{QCD}/{m_b}$, where $\Lambda_{QCD}$ is the typical hadronic scale, and $m_b$ is the bottom quark mass. Factorization of the hadronic matrix elements is proved to hold in the leading power in $\Lambda_{QCD}/{m_b}$. Within these approaches, many  decay modes have been studied, including the branching fractions and $CP$ asymmetries.

Specifically, SCET has been regarded as an effective field theory for describing the dynamics of highly energetic particles moving on the light-cone and interacting with a background field of soft quanta. SCET was put forward for describing the interactions between soft and collinear degrees of freedom in the presence of a hard interaction. In the charmless two-body nonleptonic decay $B\to M_1M_2$ with $M_2$ picking the spectator quark, the electroweak scale $m_W$ is first integrated out, and the physics running from $m_W$ to $m_b$ are resummed into Wilson coefficients within the renormalization group technique. Because the final light mesons are energetic with $E_{M_1} \simeq E_{M_2}\simeq {m_B}/{2}$, the soft spectator quark should be kicked by a hard gluon so as to become a collinear one and form an energetic light vector meson, and the typical scale taken by this hard gluon is $\Lambda_h \simeq \sqrt{m_b \Lambda_{QCD}}$. Thus, there exist three typical scales in $B\to M_1M_2$ decay:  the hard scale $m_b$, the soft scale $\Lambda_{QCD}$ set by the typical momentum of the light degrees of freedom in the heavy $B$ meson, and the collinear scale $\Lambda_h$ arising from the interaction between the collinear quark and soft modes. SCET provides us an elegant way to separate the scales in $B$ decay by matching from full QCD onto SCET in two steps. The hard degree $\mathcal{O}(m_b^2)$ is first integrated out by matching full QCD to $\mathrm{SCET_I}$, in which the generic expression for nonleptonic decay $B \to M_1 M_2$ amplitude is schematically written as
\begin{align}\label{SCET1}
\mathcal{A}&\sim T_{\zeta} \otimes \phi_{M_1}\otimes \zeta^{BM_2} +T_{J}(u,z) \otimes \phi_{M_1}(u)\otimes \zeta_J^{BM_2}(z)+...,
\end{align}
where $T_{\zeta}$ and $T_{J}$ are the perturbatively calculable Wilson coefficients. The remaining degrees of freedoms, collinear and soft, are contained in $\zeta_{(J)}^{BM_2}$ and $\phi_{M_1}$. Secondly, the fluctuations with typical off-shellness $m_b\Lambda_{QCD}$ are integrated out and one reaches  $\mathrm{SCET_{II}}$. The function $\zeta_{J}^{BM_2}$ can be further factorized into the convolution of a hard kernel (jet function) with light-cone distribution amplitudes:
\begin{align}\label{SCET2}
\zeta_J^{BM_2}\emph{}(z)=\phi_B(k_+)\otimes J(z,x,k_+)\otimes \phi_{M_2}(x).
\end{align}
The jet function $J(z,x,k_+)$ depends on physics at the intermediate scale $\Lambda_h$, so its perturbative expansion in $\alpha_s$ is not as convergent as for the $T_{\zeta}$ and $T_{J}$, which means that for the perturbative calculation of jet function $J$ might not be reliable. In Refs. \cite{Bauer:2004tj,Bauer:2005kd},  the authors pointed out that it is not mandatory to integrate out this intermediate scale from the function $\zeta_{J}$, and instead one may treat both $\zeta$ and $\zeta_{J}$ as nonperturbative parameters. These parameters  can  be obtained by fitting   the experimental data.

When discussing the charmless nonleptonic $B$ decays, an important contribution is from the intermediate charm quark loop, namely charming penguin \cite{Colangelo:1989gi}, although whether it belongs to leading power or not is still in controversy~\cite{Beneke:2009az}. In fact,   the charming penguin contains two intermediate charms. The charm quarks are moving slowly,  which means that the perturbative calculation may be questionable. Then, this contribution can be viewed as nonperturebative, whose explicit values can be fitted from data. This method was firstly applied to analyze $B\to K\pi$, $B\to KK$, $B\to \pi\pi$ decays in~\cite{Bauer:2005kd}. Subsequently, it was extended to study other charmless $B\to PP$ \cite{Williamson:2006hb} and $B\to PV$~\cite{Wang:2008rk} decays, even decays involving the isosinglet mesons. With the strong phases from the charming penguins, most branching fractions and CP asymmetries are in agreement with the experimental data. However, to our best knowledge, the global analysis of charmless $B\to VV$ decays in SCET has not been preformed. Therefore,  the purpose of this work is to fill this gap.

When the final states consist of two vector mesons, e.g. $B \to V V$, three independent polarizations  are involved. So, a complete set of polarization observables   can be experimentally measurable. In 2003, a  large transverse polarization fraction (about $50\%$) in $ B \to K^*\phi$ decays  was surprisingly  observed in Belle collaboration \cite{Aubert:2003mm, Chen:2003jfa}. Subsequently, this polarization anomaly was also confirmed  in other penguin-dominated decays, such as $B_s \to \phi \phi$ and $B_s \to K^{*0} \overline{K}^{*0}$. Similarly,  large transverse polarizations were also found in the color-suppressed tree-dominated decays, for example, the longitudinal polarization fraction of $B^0 \to \rho^0 \rho^0$ is only $f_L \approx 0.6$. These deviations from the naive power counting  $f_L \approx 1$~\cite{Korner:1979ci} have motivated a number of theoretical studies in two distinct directions. One focuses on the uncontrollable  strong interaction effects~\cite{Li:2003he,Kagan:2004uw,Beneke:2006hg,Cheng:2008gxa,Li:2004ti,Ali:2007ff,Zou:2015iwa,Ladisa:2004bp,Wang:2017hxe},  while  the other explores the   new physics effects~ \cite{Hou:2004vj, Yang:2004pm, Das:2004hq, Kim:2004wq, Zou:2005gw, Huang:2005if, Baek:2005jk, Huang:2005qb, Bao:2008hd}. In this work, we will fit 8 free parameters for $B\to VV$ decays in the SCET framework using the 35 experimental data. Then, we will calculate the branching fractions, $CP$ asymmetries and polarization observables of all charmless $B \to VV$ decays.

The rest of this paper is organized as follows.  In Sec.\ref{Sec-2}, the SCET framework for $B$ decays will be reviewed briefly. The factorization  formulas of the $B\to VV$ decay amplitudes at leading power in $1/m_b$ and leading order in $\alpha_s$ expansion are also given in this section. Numerical results for  branching fractions, {\it CP} asymmetries and polarization observables are presented  in Sec.\ref{Sec-3}. We will summarize this work at Sec.\ref{Sec-4}.

\section{$B \to V V$ decay amplitude at leading power in SCET}\label{Sec-2}
In this section, we will recapitulate the derivation of the factorization analysis at the leading power in SECT. We start from the effective weak Hamiltonian describing $b \to q (q=d,s)$ transitions, which is given as \cite{Buchalla:1995vs}
\begin{eqnarray}\label{full QCD}
\mathcal{H}_{eff}=\frac{G_F}{\sqrt 2}  \left\{ V_{ub}V^*_{uq}\left[ C_1(\mu)O_1(\mu)+C_2(\mu)O_2(\mu)\right]-V_{tb}V^*_{tq}\sum_{i=3}^{10,7\gamma,8g}C_i(\mu)O_i(\mu) \right\}+h.c. ,
\end{eqnarray}
where $G_F$ is the Fermi coupling constant, $V_{ub}V^*_{uq}$ and $V_{tb}V^*_{tq}$ are the products of the Cabibbo-Kobayashi-Maskawa (CKM) matrix elements. $O_i(\mu)$ are the local four-quark operators with the corresponding Wilson coefficients $C_i(\mu)$ at the scale $\mu$. Explicitly, the local four-quark operators are given by
\begin{align}\label{EW1}
O_1&=(\overline u_\alpha b_\beta)_{V-A}(\overline q_\beta u_\alpha)_{V-A}, \qquad ~~~~~~~~
O_2=(\overline u_\alpha b_\alpha)_{V-A}(\overline q_\beta u_\beta)_{V-A},\\
O_{3,5}&=(\overline q_\alpha b_\alpha)_{V-A}\sum_{q'}(\overline q_\beta' q_\beta')_{V\mp A}, \qquad ~~~
O_{4,6}=(\overline q_\beta b_\alpha)_{V-a}\sum_{q'}(\overline q_\alpha' q_\beta')_{V-A},\\
O_{7,9}&=\frac{3}{2}(\overline q_\alpha b_\alpha)_{V-A}\sum_{q'}e_{q'}(\overline q_\beta' q_\beta')_{V\pm A}, \quad
O_{8,10}=\frac{3}{2}(\overline q_\beta b_\alpha)_{V-A}\sum_{q'}e_{q'}(\overline q_\alpha' q_\beta')_{V\pm A},
\end{align}
and magnetic moment operators are\begin{align}
O_{7\gamma,8g}&=-\frac{m_b}{4\pi^2}\overline{q}\sigma^{\mu,\nu}\{ eF_{\mu\nu}, gG_{\mu,\nu} \}P_Rb,
\end{align}
with $(\overline q_\alpha' q_\beta')_{V \pm A} \equiv \overline q_\alpha'\gamma_{\mu}(1\pm \gamma_5) q_\beta'$ and the projection operator $P_R \equiv (1+\gamma_5)/2$. $\alpha$ and $\beta$ are the SU(3) color indices, and $q'=u, d, s, c, b$ stand for the active quarks at the scale $m_b$. For the Wilson coefficients, we will work to the leading logarithm in the naive dimensional regularization (NDR) scheme. Taking $\alpha_s(m_b)=0.119$, $\alpha_{\mathrm{em}}=1/128$ and $m_t=174.3$ GeV, we obtain the Wilson coefficients for the tree and penguin operators at scale $\mu =m_b=4.8$ GeV as\cite{Buchalla:1995vs}
\begin{align}\label{C1-6}
C_{1-6}(m_b)=\{ 1.110,-0.253, 0.011, -0.026, 0.008, -0.032 \}.
\end{align}
The Wilson coefficients for the electroweak operators are
\begin{align}\label{C7-10}
C_{7-10}(m_b)=\{ 0.09, 0.24, -10.3, 2.2 \}\times 10^{-3},
\end{align}
and   Wilson coefficients for the magnetic operators are
\begin{align}\label{C7G8G}
C_{7\gamma}(m_b)=-0.315, C_{8g}(m_b)=-0.149.
\end{align}

As aforementioned, in the charmless $B$ meson decays, there are three typical scales, hard scale $m_b$, hard-collinear scale $\Lambda_h$, and the soft scale $\Lambda_{QCD}$. These three scales would be separated by matching the effective weak Hamiltonian (\ref{full QCD}) of full QCD to their corresponding expressions in SCET. The general structure of factorization in SCET can be derived by a two-step matching from $\mathrm{QCD} \to \mathrm{SCET_I}\to \mathrm{SCET_{II}}$ as follows: the fluctuations with off-shellness $\mathcal{O}(m_b^2)$  are first integrated out to give $\mathrm{SCET_I }$ characterized by the hard-collinear scale $\Lambda_h$, which will be discussed in detail in the following subsection. Then, we integrate out the hard-collinear modes with off-shellness $\mathcal{O}(\Lambda_h^2 )$ to derive $\mathrm{SCET_{II}}$. In the whole work, we will adopt the notations as in \cite{Bauer:2004tj}, in which the recoiling meson goes in  the  $n= (1, 0, 0, 1) $ direction, while  the emitted meson moves along $\overline n=(1, 0, 0, -1)$ direction.

\subsection{Matching to $\mathrm{SCET_I}$}\label{SCETI}
Integrating out the hard scales with typical off-shellness $m_b^2$, the $\mathrm{SCET_I}$  is obtained when matching onto the full QCD theory at scale $m_b$. In order to give a complete set of the leading order contributions when matching onto $\mathrm{SCET_{II}}$, the operators in $\mathrm{SCET_I}$ should be expanded to next-to-leading order in $\lambda=\sqrt{\Lambda_{QCD}/m_b}$ as \cite{Bauer:2004tj},
\begin{eqnarray} \label{amscet}
\mathcal{H}_W=\frac{2G_F}{\sqrt{2}}\sum_{n,\overline{n}}\left[\sum_i\int [d\omega_j]^3_{j=1}c_i^{(f)}(\omega_j)Q_{if}^{(0)}(\omega_j) + \sum_i\int [d\omega_j]^4_{j=1}b_i^{(f)}(\omega_j)Q_{if}^{(1)}(\omega_j) +Q_{c\overline{c}}\right],
\end{eqnarray}
where $f=d,s$. The $Q_{if}^{(0)}$ and $Q_{if}^{(1)}$ represent leading power and next-to-leading power operators respectively. Their explicit forms  will be given  in the following. The isolated operator $Q_{c\overline{c}}$ denotes the charming penguin contribution, which stands for the long distance charm quark loop effects. In fact, a rigorous account of factorization for long-distance charm effects at leading power or a complete understanding of them order by order in a power counting expansion is not yet available~\cite{Beneke:2009az}. Therefore, the long-distance charm effects, as the main source of strong phase in SCET, can only be estimated by parameterization in SCET now. Here we will treat them  as  nonperturbative but  universal parameters.

The $\mathcal{O}(\lambda^0)$ operators $Q_{if}^{(0)}$ for $\Delta S=0$ transition are~\cite{Bauer:2004tj, Williamson:2006hb}
\begin{align} \label{LOoperator}
Q_{1d}^{(0)}&=[\overline{u}_{n,\omega_1}\overline{\slashed{n}}P_Lb_v][\overline{d}_{\overline{n},\omega_2}\slashed{n}P_Lu_{\overline{n},\omega_3}],\qquad ~~~~~~~~
Q_{2d,3d}^{(0)} =[\overline{d}_{n,\omega_1}\overline{\slashed{n}}P_Lb_v][\overline{u}_{\overline{n},\omega_2}\slashed{n}P_{L,R}u_{\overline{n},\omega_3}],  \nonumber \\
Q_{4d}^{(0)}&=[\overline{q}_{n,\omega_1}\overline{\slashed{n}}P_Lb_v][\overline{d}_{\overline{n},\omega_2}\slashed{n}P_Lq_{\overline{n},\omega_3}], \qquad ~~~~~~~~
Q_{5d,6d}^{(0)} =[\overline{d}_{n,\omega_1}\overline{\slashed{n}}P_Lb_v][\overline{q}_{\overline{n},\omega_2}\slashed{n}P_{L,R}q_{\overline{n},\omega_3}],
\end{align}
where only four quark fields are involved. As for the $\mathcal{O}(\lambda)$ operators, an additional transverse gluon ${\mathcal B}^{\perp\,\mu}_{n,\omega}$ involves, and the relevant operators are given as
\begin{eqnarray}\label{NLOoperator}
Q_{1d}^{(1)}&=\frac{-2}{m_b}[\overline{u}_{n,\omega_1}ig\slashed{\mathcal{B}}_{n,\omega_4}^{\perp}P_Lb_v] [\overline{d}_{\overline{n},\omega_2}\slashed{n}P_Lu_{\overline{n},\omega_3}],
Q_{2d,3d}^{(1)}=\frac{-2}{m_b}[\overline{d}_{n,\omega_1}ig\slashed{\mathcal{B}}_{n,\omega_4}^{\perp}P_Lb_v] [\overline{u}_{\overline{n},\omega_2}\slashed{n}P_{L,R}u_{\overline{n},\omega_3}], \nonumber \\
Q_{4d}^{(1)}&=\frac{-2}{m_b}[\overline{q}_{n,\omega_1}ig\slashed{\mathcal{B}}_{n,\omega_4}^{\perp}P_Lb_v] [\overline{d}_{\overline{n},\omega_2}\slashed{n}P_Lq_{\overline{n},\omega_3}],
Q_{5d,6d}^{(1)}=\frac{-2}{m_b}[\overline{d}_{n,\omega_1}ig\slashed{\mathcal{B}}_{n,\omega_4}^{\perp}P_Lb_v] [\overline{q}_{\overline{n},\omega_2}\slashed{n}P_{L,R}q_{\overline{n},\omega_3}], \nonumber \\
Q_{7d}^{(1)}&=\frac{-2}{m_b}[\overline{u}_{n,\omega_1}ig\mathcal{B}_{n,\omega_4}^{\perp \mu}P_Lb_v] [\overline{d}_{\overline{n},\omega_2}\slashed{n}\gamma_\mu^\perp P_Ru_{\overline{n},\omega_3}],
Q_{8d}^{(1)}=\frac{-2}{m_b}[\overline{q}_{n,\omega_1}ig\mathcal{B}_{n,\omega_4}^{\perp \mu}P_Lb_v] [\overline{d}_{\overline{n},\omega_2}\slashed{n}\gamma_\mu^\perp P_Rq_{\overline{n},\omega_3}],
\end{eqnarray}
with $Q_{is}$ obtained by replacing $d\to s$. The quark fields with subscripts $n$ and $\overline n$ are products of collinear quark fields and Wilson lines with large momenta $\omega_i$. For instance, $\overline{u}_{n,\omega}=[\overline{\xi}_n^{(u)}W_n\delta(\omega-\overline{n}\cdot \mathcal{P}^\dag)]$, where $\overline{\xi}_n$ creates a collinear up quark moving alone the $n$ direction. The definition of $ ig\mathcal{B}_{n,\omega}^{\perp\mu}$ is given as
\begin{eqnarray}
ig\mathcal{B}_{n,\omega}^{\perp\mu}=\frac{1}{(-\omega)}\{W_n^\dag[i\overline{n}\cdot D_{c,n}, iD_{n,\perp}^{\mu}]W_n\delta(\omega-\overline{\mathcal{P}}^\dag)\}.
\end{eqnarray}

Note that in eqs. (\ref{LOoperator}) and (\ref{NLOoperator}), the field products within the square brackets are color-singlet,  while the color-octet operators with $T^A\otimes T^A$ color structure in $Q_i^{(0)}$ are not listed here, because they give vanishing matrix elements of $B \to VV$ at leading power. At LO in SCET $Q_{if}^{(0,1)}$ for $i=1-6$ have scalar bilinears and give vanishing contributions to $B\to V_T V_T$ \cite{Kagan:2004uw}. Therefore only $A_{c\overline{c}}$ contribute to transverse amplitudes at LO. As for the operators $Q_{7f,8f}^{(1)}$, according to the analysis of Ref.\cite{Williamson:2006hb}, they only contribute to $B^*$ decays, and do not affect $B$ decays at all, when matching $\mathrm{SECT_I}$ onto $\mathrm{SECT_{II}}$. To further reduce the number of operators in (\ref{LOoperator}) and (\ref{NLOoperator}), $e_q \overline{q} q=\overline {u} u+\overline {c} c - \frac{1}{3}\overline {q} q $ are adopted to  eliminate  the electroweak penguin operators $O_9,O_{10}$. With the minimal choice,  the tree level matching coefficients for the relevant four-body operators $Q_{if}^{(0)}$ and five-body operators $Q_{if}^{(1)}$ in (\ref{amscet}) are
\begin{align}\label{SCETWC}
c_{1,2}^{(f)}&=\lambda_u^{(f)}[C_{1,2}+\frac{1}{N_c}C_{2,1}]-\lambda_t^{(f)}\frac{3}{2}[C_{10,9}+\frac{1}{N_c}C_{9,10}],\nonumber\\
c_{3}^{(f)}&=-\lambda_t^{(f)}\frac{3}{2}[C_{7}+\frac{1}{N_c}C_{8}],\nonumber \\
c_{4,5}^{(f)}&=-\lambda_t^{(f)}[C_{4,3}+\frac{1}{N_c}C_{3,4}-\frac{1}{2}C_{10,9}-\frac{1}{2N_c}C_{9,10}], \nonumber\\
c_{6}^{(f)}&=-\lambda_t^{(f)}[C_{5}+\frac{1}{N_c}C_{6}-\frac{1}{2}C_{7}-\frac{1}{2N_c}C_{8}];
\nonumber \\
b_{1,2}^{(f)}&=\lambda_u^{(f)}[C_{1,2}+\frac{1}{N_c}(1-\frac{m_b}{\omega_3})C_{2,1}]-\lambda_t^{(f)}\frac{3}{2}[C_{10,9}+\frac{1}{N_c}(1-\frac{m_b}{\omega_3})C_{9,10}],
\nonumber \\
b_3^{(f)}&=-\lambda_t^{(f)}\frac{3}{2}[C_{7}+\frac{1}{N_c}(1-\frac{m_b}{\omega_2})C_{8}], \nonumber\\
b_{4,5}^{(f)}&=-\lambda_t^{(f)}[C_{4,3}+\frac{1}{N_c}(1-\frac{m_b}{\omega_3})C_{3,4}]+\lambda_t^{(f)}\frac{1}{2}[C_{10,9}+\frac{1}{N_c}(1-\frac{m_b}{\omega_3})C_{9,10}],
 \nonumber\\
b_{6}^{(f)}&=-\lambda_t^{(f)}[C_{5}+\frac{1}{N_c}(1-\frac{m_b}{\omega_2})C_{6}]+\lambda_t^{(f)}\frac{1}{2}[C_{7}+\frac{1}{N_c}(1-\frac{m_b}{\omega_2})C_{8}],
\end{align}
with  $\omega_2=m_b u$, $\omega_3=-m_b \overline{u}=m_b (u-1)$.

\subsection{Matching to $\mathrm{SCET_{II}}$}
The next step is to match  $\mathrm{SCET_{I}}$ onto  $\mathrm{SCET_{II}}$ by integrating out the degrees of freedom $p^2\sim\Lambda_{QCD} m_b$. In fact, this step can be conveniently  preformed by the field redefinition, such as  $\xi_{n'}\to Y_{n'}\xi_{n'}$, $A_{n'}\to Y_{n'}A_{n'}Y_{n'}^\dag$, where $Y_{n'}$ are Wilson line of $n'\cdot A_{us}$ and $n'=n(\overline{n})$. In this way, all interactions with ultrasoft gluons are summed into Wilson line $Y_{n'}$ and only appear in the combination $Y_{n'}^\dag b_v$ in operators $Q_{if}^{(0,1)}$ (\ref{LOoperator}) and (\ref{NLOoperator}), therefore collinear sectors and anticollinear sector decouple. That is, the $\mathrm{SCET_I}$ operators can be written as
\begin{align}\label{4Q}
Q_{if}^{(0,1)}=\tilde{Q}_{if}^{n}Q_{if}^{\overline{n}}.
\end{align}
If $f=d$, the operators $\tilde{Q}_{id}^{n}$ and $Q_{id}^{\overline{n}}$ are defined as
\begin{align}
Q_{1d}^{\bar{n}(0,1)}&=\overline{d}_{\overline{n},\omega_2}\slashed{n}P_Lu_{\overline{n},\omega_3}, \qquad ~~ Q_{2d,3d}^{\bar{n}(0,1)}=\overline{u}_{\overline{n},\omega_2}\slashed{n}P_Lu_{\overline{n},\omega_3}, \nonumber \\
Q_{4d}^{\bar{n}(0,1)}&=\overline{d}_{\overline{n},\omega_2}\slashed{n}P_Lq_{\overline{n},\omega_3}, \qquad ~~ Q_{5d,6d}^{\bar{n}(0,1)}=\overline{q}_{\overline{n},\omega_2}\slashed{n}P_Lu_{\overline{q},\omega_3}, \nonumber \\
Q_{7d}^{\bar{n}(1)}&=\overline{d}_{\overline{n},\omega_2}\slashed{n}\gamma_\mu^\perp P_Ru_{\overline{n},\omega_3}, \qquad Q_{8d}^{\bar{n}(1)}=\overline{d}_{\overline{n},\omega_2}\slashed{n}\gamma_\mu^\perp P_Rq_{\overline{n},\omega_3},
\end{align}
and
\begin{eqnarray}
\tilde{Q}_{1d}^{(0)}&=\overline{u}_{n,\omega_1}\overline{\slashed{n}}P_LY_{n'}^{\dag}b_v, \qquad ~~~~ ~~~~ ~~
\tilde{Q}_{2d,3d,5d,6d}^{(0)}=\overline{d}_{n,\omega_1}\overline{\slashed{n}}P_LY_{n'}^{\dag}b_v, \nonumber \\
\tilde{Q}_{4d}^{(0)}&=\overline{q}_{n,\omega_1}\overline{\slashed{n}}P_LY_{n'}^{\dag}b_v, \qquad ~~~~ ~~~~ ~~ \tilde{Q}_{1d}^{(1)}=\frac{-2}{m_b}[\overline{u}_{n,\omega_1}ig\slashed{\mathcal{B}}_{n,\omega_4}^{\perp}P_LY_{n'}^{\dag}b_v], \nonumber \\
\tilde{Q}_{4d}^{(1)}&=\frac{-2}{m_b}[\overline{q}_{n,\omega_1}ig\slashed{\mathcal{B}}_{n,\omega_4}^{\perp}P_LY_{n'}^{\dag}b_v], ~~ \tilde{Q}_{2d,3d,5d,6d}^{(1)}=\frac{-2}{m_b}[\overline{d}_{n,\omega_1}ig\slashed{\mathcal{B}}_{n,\omega_4}^{\perp}P_LY_{n'}^{\dag}b_v], \nonumber \\
\tilde{Q}_{7d}^{(1)}&=\frac{-2}{m_b}[\overline{u}_{n,\omega_1}ig\mathcal{B}_{n,\omega_4}^{\perp \mu}P_LY_{n'}^{\dag}b_v], \quad
\tilde{Q}_{8d}^{(1)}=\frac{-2}{m_b}[\overline{q}_{n,\omega_1}ig\mathcal{B}_{n,\omega_4}^{\perp \mu}P_LY_{n'}^{\dag}b_v].
\end{eqnarray}
The $Q_{if}^{\overline{n}}$ operators relate to the emitted meson, which can be factorized out directly, 
\begin{align}
\langle V_{L} |Q_{if}^{\overline{n}}|0\rangle=m_Bf_V\phi_{V_L}.
\end{align}

In the following we concentrate on the remaining part $\tilde Q_{if}^{n}$, which relates to the $B\to V$ form factors. To describe the nonlocal interaction between $\tilde{Q}_{if}^{n}$, the spectator quark and the changed gluons, the form factors contain  the  time-ordered products. At leading power,
\begin{eqnarray}
&&T_1=\int d^4yd^4y'{\rm T}\left [\tilde{Q}_{if}^{(0)}(0), i\mathcal{L}_{\xi_nq}^{(1)}(y), i\mathcal{L}_{\xi_n\xi_n}^{(1)}(y')+i\mathcal{L}_{cg}^{(1)}(y')] +\int d^4yT[\tilde{Q}_{if}^{(0)}(0),i\mathcal{L}_{\xi_nq}^{(1,2)}(y)\right ], \nonumber \\
&&T_2  =\int d^4y{\rm T}\left [\tilde{Q}_{if}^{(1)}(0),i\mathcal{L}_{\xi_nq}^{(1)}(y)\right ],
\end{eqnarray}
this are obtained by inserting the next-to-leading power Lagrangian $\mathcal{L}_{\xi_n q}^{(1)}=\overline{q}_{us}Yig\slashed{\mathcal{B}}_n^\perp W^\dag\xi_n+h.c.$ \cite{Beneke:2002ph}, which boosts the soft spectator quark to the collinear quark. The form of other $\mathcal{L}'s$ can be found in \cite{Bauer:2002aj}. Then, we will match these products $T_{1,2}$ in $\mathrm{SCET}_I$ onto $\mathrm{SCET}_{II}$. As mentioned before, only $A_{c\overline{c}}$ contributes to the transversely polarized  states, so $\langle V_T|T_{1,2}|B\rangle=0 $. For the longitudinal state $V_L$, due to the the end-point singularity, $T_{1}$ cannot be further factorized, then the nonzero matrix element can be parameterized as
\begin{align}
\langle V_L |T_1|B\rangle=m_B\zeta^{BV_L}.
\end{align}

Nevertheless, the matrix element $\langle V_L |T_2|B\rangle$ can be further factorized into a convolution of light-cone distribution amplitudes (LCDAs) and jet function $J$. Alternatively, instead of integrating out $\sqrt{\Lambda m_B}$ scale, we directly parameterize the matrix elements induced by $T_2$ as
\begin{align}\label{zetaJ}
\langle V_L |T_2|B\rangle=m_B\zeta_J^{BV_L}.
\end{align}

\subsection{Factorization  formulas for $B \to VV$}
With the  above definitions, we can write the factorization  amplitude for the $B\to V_L V_L$  as
\begin{align}\label{ALL}
\mathcal{A}_L(B\to V_1V_2)=&\frac{G_F}{\sqrt{2}}m_B^2\Big\{f_{V_1} \int du \phi_{V_1}(u)\int dzT_{1J}(u,z)\zeta_J^{BV_2}(z)+f_{V_1}\int du\phi_{V_1}(u)T_{1\zeta}(u)\zeta^{BV_2}   \nonumber \\
&+(1\leftrightarrow 2) +\lambda_c^{(f)}A_{cc}^{V_1V_2}  \Big\}.
\end{align}
At the leading order in $\alpha_s(m_b)$, the hard kernels $T_{1\zeta,2\zeta}$ are constants and $T_{1J,2J}(u,z)$ are functions of $u$ only, then the above amplitude can be further simplified as
\begin{align}
\mathcal{A}_L&=\frac{G_F}{\sqrt{2}}m_B^2\Big\{ f_{V_1}\zeta_J^{BV_2}\int du\phi_{V_1}(u)T_{1J}(u)+f_{V_1}\zeta^{BV_2}T_{1\zeta}+ (1\leftrightarrow 2)+\lambda_c^{(f)}A_{cc L}^{V_1V_2} \Big\},
\end{align}
where
\begin{eqnarray}
\zeta_J^{BV_1}=\int \zeta_J^{BV_1}(z)dz.
\end{eqnarray}
For the transverse polarized parts, since at LO only the charming penguins $A_{c\overline{c}}$ contribute, the amplitudes can be parameterized as
\begin{align}
\mathcal{A}_\parallel=\frac{G_F}{\sqrt{2}}m_B^2\lambda_c^{(f)}A_{cc\parallel}^{V_1V_2}, \qquad \mathcal{A}_\perp=\frac{G_F}{\sqrt{2}}m_B^2\lambda_c^{(f)}A_{cc\perp}^{V_1V_2}.
\end{align}

In order to reduce the independent inputs and reinforce the predictive power, we will employ the SU(3) symmetry for $B$ to light form factors and charming penguins. So, there are only two form factors left, 
\begin{align}
\zeta_{(J)}\equiv \zeta_{(J)}^{B\rho}=\zeta_{(J)}^{BK^*}=\zeta_{(J)}^{B\omega}=\zeta_{(J)}^{B_sK^*}=\zeta_{(J)}^{B_s\phi}.
\end{align}

Now, the left unknown non-perturbative parameters are   form factors $\zeta$ and $\zeta_J$ and three complex charming penguins  $A_{cc(L, \parallel, \perp)}$. So, we have totally 8 parameters, which can be extracted from experimental data. For each decay mode, the hard coefficients $T_{i\zeta}$ are given in in Table. \ref{Tid} and Table. \ref{Tis} for $\Delta S=0$ transitions and $\Delta S=1$ transitions, respectively. The last column are the coefficients of $A_{cc}$. The coefficients $T_{iJ}$ can be easily obtained by replacement of $c^f_i \to b^f_i$ in $T_{i\zeta}$.
\begin{table}[h]
\caption{Hard kernels $T_{i\zeta}$ and the coefficients of charming penguins $A_{cc}$ for $\Delta S=0$ decays. The superscripts $c_i^{(d)}$ are not displayed for brevity.}\label{Tid}
\centering
\begin{tabular}{ccccc}
\cmidrule(lr){1-5}\morecmidrules\cmidrule(lr){1-5}
 &Mode & $T_{1\zeta}$ & $T_{2\zeta}$ & coeff. of $A_{cc}$ \\
\hline
$B^-$&$\rho^0\rho^-$ & $\frac{1}{\sqrt{2}}(c_2+c_3-c_4)$ &$\frac{1}{\sqrt{2}}(c_1+c_4)$  &$0 $ \\
&$\rho^-\omega$ & $\frac{1}{\sqrt{2}}(c_1+c_4)$ & $\frac{1}{\sqrt{2}}(c_2+c_3+c_4)+\sqrt{2}(c_5+c_6)$ & $\sqrt{2}$\\
&$\rho^-\phi$  & 0 &$c_5+c_6$ & 0 \\
&$K^{*0}K^{*-}$ & $c_4$ & 0 &1 \\
\hline
$\overline{B}^0$&$\rho^0\rho^0$ & $\frac{1}{2}(-c_2-c_3+c_4)$ & $\frac{1}{2}(-c_2-c_3+c_4)$ & 1 \\
&$\rho^+\rho^-$ & $0$ & $c_1+c_4$ & 1 \\
&$\rho^0\omega$ & $\frac{1}{2}(c_2+c_3-c_4)$ & $-\frac{1}{2}(c_2+c_3+c_4)-c_5-c_6$ & -1 \\
&$\omega\omega$ & $\frac{1}{2}(c_2+c_3+c_4)+c_5+c_6$ & $\frac{1}{2}(c_2+c_3+c_4)+c_5+c_6$ & 1 \\
&$\rho^0\phi$ & $0$ & $-\frac{1}{\sqrt{2}}(c_5+c_6)$ & 0 \\
&$\omega\phi$ & $0$ & $\frac{1}{\sqrt{2}}(c_5+c_6)$ & 0 \\
&$K^{*0}\overline{K}^{*0}$ & $c_4$ & $0$ & 1 \\
\hline
$B_s$&$K^{*0}\rho^0$ & $\frac{1}{\sqrt{2}}(c_2+c_3-c_4)$ & $0$ & $-\frac{1}{\sqrt{2}}$ \\
&$K^{*0}\omega$ & $0$ & $\frac{1}{\sqrt{2}}(c_2+c_3+c_4)+\sqrt{2}(c_5+c_6)$ & $\frac{1}{\sqrt{2}}$ \\
&$K^{*+}\rho^-$ & $c_1+c_4$ & $0$ & 1 \\
&$K^{*0}\phi$ & $c_4$ & $c_5+c_6$ & 1 \\
\cmidrule(lr){1-5}\morecmidrules\cmidrule(lr){1-5}
\end{tabular}
\end{table}
\begin{table}[h]
\caption{Hard kernels and Wilson coefficients of charming penguins for $\Delta S=1$ decays. The superscripts $c_i^{(s)}$ are not displayed for brevity.}\label{Tis}
\centering
\begin{tabular}{ccccc}
\cmidrule(lr){1-5}\morecmidrules\cmidrule(lr){1-5}
&Mode & $T_{1\zeta}$ & $T_{2\zeta}$ & coeff. of $A_{cc}$ \\
\hline
$B^-$&$\rho^0K^{*-}$ & $\frac{1}{\sqrt{2}}(c_2+c_3)$ &$\frac{1}{\sqrt{2}}(c_1+c_4)$  &$\frac{1}{\sqrt{2}} $ \\
&$\rho^-\overline{K}^{*0}$ & $0$ & $c_4$ & $1$\\
&$K^{*-}\omega$  & $\frac{1}{\sqrt{2}}(c_1+c_4)$ &$\frac{1}{\sqrt{2}}(c_2+c_3)+\sqrt{2}(c_5+c_6)$ & $\frac{1}{\sqrt{2}}$ \\
&$K^{*-}\phi$ & $0$ & $c_4+c_5+c_6$ &1 \\
\hline
$\overline{B}^0$&$\rho^0\overline{K}^{*0}$ & $\frac{1}{2}(c_2+c_3)$ & $-\frac{1}{\sqrt{2}}c_4$ & $-\frac{1}{\sqrt{2}}$ \\
&$\rho^+K^{*-}$ & $0$ & $c_1+c_4$ & 1 \\
&$\overline{K}^{*0}\omega$ & $\frac{1}{\sqrt{2}}c_4$ & $\frac{1}{2}(c_2+c_3)+\sqrt{2}(c_5+c_6)$ & $\frac{1}{\sqrt{2}}$ \\
&$\overline{K}^{*0}\phi$ & $0$ & $c_4+c_5+c_6$ & $\frac{1}{2}$ \\
\hline
$B_s$&$\rho^0\phi$ & $\frac{1}{\sqrt{2}}(c_2+c_3)$ & $0$ & $0$ \\
&$\omega\phi$ & $\frac{1}{\sqrt{2}}(c_2+c_3)+\sqrt{2}(c_5+c_6)$ & $0$ & $0$ \\
&$\phi\phi$ & $c_4+c_5+c_6$ & $c_4+c_5+c_6$ & $2$ \\
&$K^{*0}\overline{K}^{*0}$ & $0$ & $c_4$ & 1 \\
&$K^{*+}K^{*-}$ & $0$ & $c_1+c_4$ & 1 \\
\cmidrule(lr){1-5}\morecmidrules\cmidrule(lr){1-5}
\end{tabular}
\end{table}
\section{Numerical analysis}\label{Sec-3}
At the beginning of this section, we shall list the input parameters used in  our numerical calculations. The CKM matrix elements, in the form of the Wolfenstein parametrization, are given as~\cite{cpc2016}
\begin{eqnarray}
\lambda = 0.22537 \pm 0.00061,\,\,\, A = 0.814^{+0.023}_{-0.024},\,\,\,
\overline \rho = 0.117 \pm 0.021,\,\,\, \overline \eta = 0.353\pm 0.013.
\end{eqnarray}
The decay constants of the vector mesons  are used as \cite{Wang:2017hxe}
\begin{align}\label{decayconstant}
f_\rho=0.213~{\rm GeV}, \quad f_\omega=0.192 {\rm GeV}, \quad f_\phi=0.225 {\rm GeV}, \quad f_{K^*}=0.220 {\rm GeV}.
\end{align}
For the inverse moment of light-cone distribution amplitudes of vector mesons, $\langle x^{-1}\rangle=3(1-a_1+a_2)$, $\langle \bar{x}^{-1}\rangle =3(1+a_1+a_2)$, $a_{1,2}$ are the Gegenbauer moments, the values we adopted are from the QCD sum rules \cite{Ball:2007rt}.

Because $B\to VV$ decay contains three kinds of polarization contributions, the observables are more complicated  than that of $B\to P P, PV$ decays. A typical set of observables consists of the branching fraction, two observables formed by the three polarization fractions $f_L$, $f_{\parallel}$, $f_{\perp}$, and two relative phases of transversal amplitudes, $\phi_{\parallel},\phi_{\perp}$. The decay width is expressed as
\begin{align}\label{Gamma}
\Gamma=\frac{|\mathbf{p}|}{8\pi m_B^2}(|\mathcal{A}_L|^2+|\mathcal{A}_\parallel|^2+|\mathcal{A}_\perp|^2),
\end{align}
where $\mathbf{p}$ is the 3-dimension momentum of each meson in the final state in the center-of-mass frame of $B$ meson.   $\mathcal{A}_L$, $\mathcal{A}_{\parallel}$ and $\mathcal{A}_{\perp}$ are three polarized decay amplitudes of $B\to VV$.  The definitions of the polarization fractions $f_{L,\parallel,\perp}$, and the relative phases $\phi_{\parallel,\perp}$ are given as:
\begin{align}\label{fL}
f_{L,\parallel,\perp}=\frac{|\mathcal{A}_{L,\parallel,\perp}|^2}{|\mathcal{A}_L|^2+|\mathcal{A}_\parallel|^2+|\mathcal{A}_\perp|^2}, \quad \phi_{\parallel,\perp}=\text{Arg}\frac{\mathcal{A}_{\parallel,\perp}}{\mathcal{A}_L}.
\end{align}
Combined with the {\it CP}-conjugate decay, the direct $CP$ asymmetry is defined as
\begin{eqnarray}
A_{CP}= \frac{\sum_{h}^{}(|\overline {\mathcal{A}}_h|^2-|\mathcal{A}_h|^2)}{\sum_{h}^{}(|\overline{ \mathcal{A}}_h|^2+|\mathcal{A}_h|^2)},
\end{eqnarray}
with $h=L,\parallel,\perp$. $\overline{\mathcal{A}}_h$ represents three polarization amplitudes of $\overline {B}\to VV$. Correspondingly, the $CP$ asymmetries of polarization fractions $f$ and relative phases $\phi$ are defined as,
\begin{align}\label{f_ACP}
A_{CP}^0=\frac{\overline{f}_L-f_L}{\overline{f}_L+f_L}, \quad A_{CP}^\perp=\frac{\overline{f}_\perp-f_\perp}{\overline{f}_\perp+f_\perp}, \quad
\Delta\phi_\parallel=\frac{\overline{\phi}_\parallel-\phi_\parallel}{2}, \quad \Delta\phi_\perp=\frac{\overline{\phi}_\perp-\phi_\perp}{2},
\end{align}
where $\overline{f}_{L,\perp}$ can also be defined as eq.(\ref{fL}).

Next, we will use the minimal $\chi^2$ fit method to determine the $8$ unknown nonperturbative parameters. From Particle Data Group (PDG) \cite{cpc2016}, we collect 35 observables of $B^{\pm/0}\to V V$ with the statistical significance  no less than $3\sigma$. In this way, we obtain the best-fitted values together with the corresponding 1$\sigma$ uncertainties
\begin{align}\label{parameters}
\zeta&=(33.3\pm1.6)\times 10^{-2}, \quad \zeta_J=(1.6\pm1.0)\times10^{-2},\nonumber\\
|A_{ccL}|&=(38.1\pm1.1)\times10^{-4},\quad \text{arg}[A_{ccL}]=-0.29\pm0.11, \nonumber \\
|A_{cc\parallel}|&=(18.8\pm0.8)\times10^{-4},\quad \text{arg}[A_{cc\parallel}]=1.98\pm0.18,\nonumber\\
|A_{cc\perp}|&=(17.1\pm0.7)\times10^{-4},\quad \text{arg}[A_{cc\perp}]=2.11\pm0.18,
\end{align}
with $\chi^2/d.o.f=67.1/(35-8)=2.5$.

One can obtain the $B\to V$ form factor $A_0$ at tree level:
\begin{align}
A_0^{B\to V}=\zeta+\zeta_J=0.349\pm0.019.
\end{align}
This is consistent with result of the light-cone sum rules \cite{Straub:2015ica}, but is much larger than the result $0.232\pm0.037$ fitted from $B\to PV$ modes \cite{Wang:2008rk}. From the solution (\ref{parameters}), we also note $\zeta\gg\zeta_J$, which differs from the case in $B\to PP$ \cite{Bauer:2005kd,Williamson:2006hb} where $\zeta$ and $\zeta_J$ are comparative. It is known that for the color-suppressed decay $B\to \pi^0\pi^0$ the theoretical predictions based on the QCD are smaller than the experimental data, which is called the ``$\pi\pi$ puzzle''. In oder to accommodate the unexpected large branching fractions, a larger $\zeta_J$ is needed in   $B \to PP$ \cite{Bauer:2005kd}. On the contrary, for $B\to \rho^0\rho^0$ decay, the experimental data indicates that the color-suppressed contribution is not as significant as  in $B\to \pi^0\pi^0$, so we  obtain a small  $\zeta_J$  here. In fact, this pattern is also confirmed in the factorization-assisted topological-amplitude approach \cite{Wang:2017hxe, Zhou:2016jkv}. In eq.(\ref{parameters}), $|A_{cc\parallel} |\approx |A_{cc\perp}|$  means that the positive helicity amplitude of charming penguin is negligible. It is compatible with the argument that the negative helicity amplitude of penguin annihilation diagram is chiral enhanced while the positive helicity amplitude is power suppressed \cite{Kagan:2004uw}.

Using the best fitted nonperturbative parameters (\ref{parameters}), we calculated  the branching fractions, direct {\it CP} asymmetries and polarization observables of 28 $B\to VV$ decays, with the numerical results listed in Table.\ref{largefl}-\ref{bqother}. For comparison,  the experimental measurements and the results from the PQCD \cite{Zou:2015iwa} and QCDF \cite{Cheng:2008gxa} are also presented. In our calculations, several approximations have been adopted, which may lead to considerable  uncertainties. For instance, in the $\chi^2$ fitting, $\rm SU(3)$ symmetry has been adopted, whose breaking can reach up to $30\%$. Furthermore, the  analysis here is at leading power and leading order in $\alpha_s$ expansion,  and the high power and high order corrections have not been included. In order to demonstrate the uncertainties induced by these approximations, we vary the magnitudes of the charming penguins by $20\%$ and the phases by $20^\circ$. These are referred to the first kind of uncertainties. The second type of errors is from statistical uncertainties of the experimental data in the fitting. The third kind of errors is  from the  uncertainties (order $5 \%$ ) in the decay constants and distribution amplitudes of vector mesons. The final errors are  combined  in quadrature due to the limit space.

For the charmless $\overline{B}\to VV$ decays, as stated previously, there are three helicity amplitudes, $\mathcal{A}^{0,\pm}$. In the light of the naive power counting of helicity flip, $\mathcal{A}^+$ and $\mathcal{A}^-$ require one and two helicity-flips respect to $\mathcal{A}^0$, respectively. So we obtain a hierarchy of helicity amplitudes
\begin{eqnarray}\label{naive fL}
{\cal A}^0:{\cal A}^-:{\cal A}^+=1:\frac{\Lambda_{QCD}}{m_b}:\left(\frac{\Lambda_{QCD}}{m_b}\right)^2,
\end{eqnarray}
which is a consequence of the helicity conservation. This hierarchy is also obtained from the  naive factorization analysis \cite{Beneke:2006hg}. Alternatively, we adopt the convention $\mathcal{A}_L=\mathcal{A}^0$, $\mathcal{A}_\parallel= (\mathcal{A}^+ +\mathcal{A}^-)/\sqrt{2}$, $\mathcal{A}_\perp=(\mathcal{A}^--\mathcal{A}^+)/\sqrt{2}$. Because the positive-helicity amplitude $\mathcal{A}^+$ is power-suppressed relative to the negative one as showing in (\ref{naive fL}), $\mathcal{A}^+$  is set to be zero approximatively in the absence of any consistent calculation of this power correction. In \cite{Beneke:2006hg}, the results indicat that the corrections to positive-helicity amplitude from weak annihilation contribution are very small, even for penguin-dominated decays. So, in this approximation, the two transverse polarization degrees of freedom have the  same power $\mathcal{A}_L:\mathcal{A}_\parallel : \mathcal{A}_\perp\simeq1:\lambda^2:\lambda^2$ and $\phi_{\parallel} \simeq \phi_{\perp}$,  which can been seen in Table. \ref{largefl}.

From the naive analysis, as shown in eq. (\ref{naive fL}), the longitudinal amplitude is dominant, whose polarization fraction $f_L$ is expected to be close to unity. However, as mentioned before, experimental measurements of $f_L$ for penguin-dominated decays, such as decay modes list in Table.~\ref{largefl}, are in conflict with this expectation. In QCDF \cite{Beneke:2006hg,Cheng:2008gxa}, the large transverse polarizations come from (i) the next-to-leading order  $\alpha_s$ corrections, which will  enhance (reduce) the transverse (longitudinal) amplitudes; (ii) the large penguin annihilations,  which were estimated by introducing non-perturbative parameters that usually cannot be computed rigorously. In PQCD approach \cite{Zou:2015iwa}, the unexpected large transverse polarization fractions were led by the penguin annihilation and the hard-scattering emission diagrams, especially the $(S+P)(S-P)$ operators.  In this work, the  large transverse polarizations arise from the nonperturbative charming penguins. From eq.~(\ref{parameters}), it is found  that the contributions form the transverse charming penguins are at the same order as the longitudinal ones, therefore the transverse polarizations in the penguin-dominated decays can be enhanced  and the  $f_L$ can be reduced to about $50\%$.

We now discuss  some specific phenomenons of the $B \to V V$ decays. First of all,  in Table.\ref{largefl}, we list our results of the well-measured penguin-dominant channels, $B^0(B^+)\to K^*(K^{*+})\phi$, $B_s\to \overline{K}^{*0}\phi$, $B_s\to\overline{K}^{*0}K^{*0}$ and $B_s\to\phi\phi$, as well as the experimental data. It is obvious that our results are in good agreement with the available experimental data except  the branching fractions of $B_s\to \overline{K}^{*0} K^{*0}$ and $B_s\to \overline{K}^{*0}\phi$ decays. The branching fractions we calculated  are much smaller than the data. It should be emphasized that, the QCDF predictions for the $B_s\to \overline{K}^{*0} K^{*0}$ and $B_s\to \overline{K}^{*0}\phi$ branching fractions are $(6.6_{-2.2}^{+2.2})\times 10^{-6}$ and $(0.37_{-0.21}^{+0.25})\times 10^{-6}$, respectively. Similarly, in PQCD, the branching fractions are predicted to be $(5.4_{-2.4}^{+3.0})\times 10^{-6}$ and $(0.39_{-0.19}^{+0.20})\times 10^{-6}$. All three theoretical predictions agree, and we then suggest the experimentalists study these two decays   in future, so as to test the SM. As  for the decays $\overline{B}_s^0 \to K^{*0}\phi$, $\overline{B}_s^0 \to \overline{K}^{*0}K^{*0}$ and $\overline{B}_s^0 \to \phi\phi$  that are induced only by the penguin operators, there is no contribution from the tree  operators, so that their direct {\it CP} asymmetries are zero in  PQCD. In QCDF, with the small contributions from the quark loops,  the direct {\it CP} asymmetries were predicted to be $(-9_{-6}^{+5})\%$, $(0.4_{-0.6}^{+1.0})\%$ and $(0.2_{-0.4}^{+0.6})\%$ respectively,  which are comparable with results of SCET (see Table.\ref{largefl}) but with a wrong sign for the central  value. The future experiment data in the ongoing LHCb and the forthcoming Belle-II can  clarify  these disagreements.

For the decays $\overline{B}^0\to K^{*+}K^{*-}$, $\overline{B}^0\to\phi\phi$, $\overline{B}_s^0\to \rho^0\rho^0$, $\overline{B}_s^0\to \rho^-\rho^+$, $\overline{B}_s^0\to \omega\rho^0$ and  $\overline{B}_s^0\to \omega\omega$, they occur only through annihilation diagrams that belong to next-leading power in SCET scheme, so that  we did not study them in the current work. The predicted observables of other remaining $B_{u,d,s} \to VV$ decays are presented in Tables.~\ref{bqbr}-\ref{bqother}.

\begin{table*}[!htb]
\centering
\caption{Branching fractions, the fractions of the longitudinal polarization $f_{L}$ and
the  transverse polarizations $f_{\perp}$, relative phases, and the $CP$ asymmetry parameters $A^{0}_{CP}$ and $A^{\perp}_{CP}$ of decays
$B \to  K^{*0}\phi$, $B^-\to K^{*-}\phi$
 $B_s \to \overline{K}^{*0}\phi$, $B_s\to \phi\phi$ and $B_s\to \overline{K}^{*0}K^{*0}$.}
 \vspace{0.cm}
\begin{tabular}[t]{lcccccccccc}
\hline\hline
 Modes& $Br(10^{-6})$ &$f_{L}$(\%) &$f_{\perp}$ (\%)& $\phi_{\parallel}$(rad)& $\phi_{\perp}$(rad)\\
 \hline
{$\overline{B}^0\to \overline{K}^{*0}\phi$} &{$9.14\pm3.14$} &{$51.0\pm16.4$} &{$22.2\pm9.9$} &{$2.41\pm0.62$} &{$2.54\pm0.62$}\\
\vspace{0.13cm}
{$Exp$} &{$10.0\pm0.5$} &{$49.7\pm1.7$} &{$22.4\pm1.5$} &{$2.43\pm0.11$} &{$2.53\pm0.09$}\\
\hline
 {$B^-\to K^{*-}\phi$}&  {$9.86\pm3.39$} &  {$51.0\pm16.4$} & {$22.2\pm9.9$}& {$2.41\pm0.62$}& {$2.54\pm0.62$}\\
\vspace{0.13cm}
 {$Exp$}             &   {$10.0\pm2.0$} &  {$50\pm5$} & {$20\pm5$}& {$2.34\pm0.18$}& {$2.58\pm0.17$}\\
\vspace{0.13cm}
 {$\overline{B}_s^0\to \phi\phi$}  &  {$19.0\pm6.5$} &  {$51.0\pm16.4$} & {$22.2\pm9.9$}& {$2.41\pm0.62$}& {$2.54\pm0.62$}\\
\vspace{0.13cm}
 {$Exp$}                      &  {$19.3\pm3.1$} &  {$36.2\pm1.4$} &  {$30.9\pm1.5$ }& {$2.55\pm0.11$}& {$2.67\pm0.23$}\\
\vspace{0.13cm}
 {$\overline{B}_s^0\to K^{*0}\phi$}  &  {$0.56\pm0.19$} &  {$54.6\pm15.0$}& {$20.5\pm9.1$}& {$2.37\pm0.59$}& {$2.50\pm0.59$}\\
\vspace{0.13cm}
 {$Exp$}                      &  {$1.13\pm0.30$} &  {$51\pm17$}      & {$28\pm11$}   & {$1.75\pm0.58\pm0.30$}& { }\\
\vspace{0.13cm}
 {$\overline{B}_s^0\to K^{*0}\overline{K}^{*0}$} &  {$8.60\pm3.07$} &  {$44.9\pm18.3$} & {$24.9\pm11.1$} & {$2.47\pm0.67$}& {$2.60\pm0.67$}\\
\vspace{0.3cm}
 {$Exp$} &  {$28\pm7$} &  {$31\pm13$} &  {$38\pm11$} & {$ $}& {$ $}\\
\hline
&$A^{dir}_{CP}$(\%)&$A^{0}_{CP}$(\%)&$A^{\perp}_{CP}$(\%)&$\Delta\phi_{\parallel}(10^{-2}\text{rad})$&$\Delta\phi_{\perp}(10^{-2}\text{rad})$&\\
\hline
\vspace{0.13cm}
 {$\overline{B}^0\to \overline{K}^{*0}\phi$}& {$-0.39\pm0.44$} & {$-0.38\pm0.45$} & {$0.39\pm0.44$}& {$-0.85\pm0.35$}& {$-0.85\pm0.35$}\\
\vspace{0.13cm}
 {$Exp$}                             &{$0\pm4$} & {$-0.7\pm3.0$} & {$-2\pm6$}& {$5\pm5$}& {$8\pm5$}\\
\vspace{0.13cm}
 {$B^-\to K^{*-}\phi$}&  {$-0.39\pm0.44$} &  {$-0.38\pm0.45$} & {$0.39\pm0.44$}& {$-0.85\pm0.35$}& {$-0.85\pm0.35$}\\
\vspace{0.13cm}
 {$Exp$}             &   {$-1\pm8$}      &  {$17\pm11$} & {$22\pm25$}         & {$7\pm21$}& {$19\pm21$}\\
\vspace{0.13cm}
 {$\overline{B}_s^0\to \phi\phi$}  &  {$-0.39\pm0.44$} &  {$-0.38\pm0.45$} & {$0.39\pm0.44$}& {$-0.85\pm0.35$}& {$-0.85\pm0.35$}\\
\vspace{0.13cm}
 {$\overline{B}_s^0\to K^{*0}\phi$}  &  {$6.61\pm7.56$} &  {$5.48\pm6.70$}& {$-6.61\pm7.56$}& {$14.6\pm5.3$}& {$14.6\pm5.3$}\\
\vspace{0.3cm}
 {$\overline{B}_s^0\to K^{*0}\overline{K}^{*0}$} &  {$-0.56\pm0.61$} &  {$-0.68\pm0.77$} &  {$0.56\pm0.61$} & {$-1.17\pm0.60$}& {$-1.17\pm0.60$}\\
 \hline\hline
\end{tabular}\label{largefl}
\end{table*}

\begin{table*}[tbh!]
\centering
\caption{Branching fractions (in units of $10^{-6}$) of  $ B_{(s)} \to V V$ decays.}
\begin{tabular}{l l l l l }
\hline
\hline
 Channel  & Exp. & SCET & PQCD & QCDF    \\
\hline
$B^-\to\rho^-\rho^0$       &$24.0\pm 1.9$    &$22.1\pm3.7$  &$13.5_{-4.1}^{+5.1}$ &$20.0_{-2.1}^{+4.5}$\\
$B^-\to\rho^-\omega$       &$15.9\pm 2.1$    &$19.2\pm3.1$  &$12.1_{-3.7}^{+4.5}$ &$16.9_{-1.8}^{+3.6}$ \\
$B^-\to K^{*0}K^{*-}$      &$1.2\pm 0.5$    &$0.52\pm0.18$  &$0.56_{-0.23}^{+0.26}$ &$0.6_{-0.3}^{+0.3}$ \\
$B^-\to\rho^-\phi$         &$<3.0$        &$0.005\pm0.001$  &$0.028_{-0.013}^{+0.015}$ &$ $             \\
$B^-\to\rho^0K^{*-}$        &$4.6\pm 1.1$   &$4.64\pm1.37$  &$6.1_{-2.4}^{+2.8}$ &$5.5_{-2.5}^{+1.4}$ \\
$B^-\to\rho^-\overline{K}^{*0}$  &$9.2\pm 1.5$   &$8.93\pm3.18$  &$9.9_{-4.1}^{+4.7}$ &$9.2_{-5.5}^{+3.8}$ \\
$B^-\to K^{*-}\omega$       &$<7.4$         &$5.56\pm1.60$  &$4.0_{-1.6}^{+2.2}$ &$3.0_{-1.5}^{+2.5}$ \\
\hline
$\overline{B}^0\to \rho^{+}\rho^{-}$  &$28.3\pm2.12$  &$27.7\pm4.1$  &$26.0_{-8.3}^{+10.3}$ &$25.5_{-3.0}^{+1.9}$ \\
$\overline{B}^0\to \rho^{0}\rho^{0}$  &$0.97\pm0.24$  &$1.00\pm0.29$  &$0.27_{-0.10}^{+0.12}$ &$0.9_{-0.45}^{+1.9}$ \\
$\overline{B}^0\to \rho^{0}\omega$     &$<1.6$        &$0.59\pm0.19$  &$0.40_{-0.14}^{+0.17}$ &$0.08_{-0.02}^{+0.36}$ \\
$\overline{B}^0\to \omega\omega$       &$1.2\pm0.4$   &$0.39\pm0.13$  &$0.50_{-0.20}^{+0.23}$ &$0.7_{-0.4}^{+1.1}$ \\
$\overline{B}^0\to K^{*0}\overline{K}^{*0}$   &            &$0.48\pm0.16$  &$0.34_{-0.15}^{+0.13}$ &$0.6_{-0.3}^{+0.2}$ \\
$\overline{B}^0\to \rho^{0}\phi$         &$<0.33$     &$\approx0.002$  &$0.013_{-0.006}^{+0.007}$ &$ $             \\
$\overline{B}^0\to \omega\phi$           &$<0.7$      &$\approx0.002$  &$0.010_{-0.004}^{+0.005}$ &$ $             \\
$\overline{B}^0\to \rho^{0}\overline{K}^{*0}$  &$3.9\pm1.3$  &$5.87\pm1.87$  &$3.3_{-1.4}^{+1.7}$ &$4.6_{-3.6}^{+3.6}$ \\
$\overline{B}^0\to \rho^{+}K^{*-}$        &$10.3\pm2.6$  &$10.6\pm3.2$  &$8.4_{-3.5}^{+3.8}$ &$8.9_{-5.6}^{+4.9}$ \\
$\overline{B}^0\to \overline{K}^{*0}\omega$     &$2.0\pm0.5$  &$3.82\pm1.39$  &$4.7_{-2.0}^{+2.6}$ &$2.5_{-1.6}^{+2.5}$ \\
\hline
$\overline{B}_s^0\to K^{*+}\rho^-$       &$ $    &$28.1\pm4.2$  &$24.0_{-9.1}^{+11.0}$  &$21.6_{-3.2}^{+1.6}$\  \\
$\overline{B}_s^0\to K^{*0}\rho^0$     &$<767$    &$1.04\pm0.30$    &$0.40_{-0.17}^{+0.22}$  &$1.3_{-0.7}^{+2.6}$\\
$\overline{B}_s^0\to K^{*0}\omega$     &$ $    &$0.41\pm0.14$    &$0.35_{-0.18}^{+0.19}$  &$1.1_{-0.6}^{+2.0}$\\
$\overline{B}_s^0\to K^{*-}K^{*+}$     &$ $    &$11.0\pm3.3$    &$5.4_{-2.3}^{+3.3}$  &$7.6_{-2.1}^{+2.5}$\\
$\overline{B}_s^0\to \phi\rho^0$       &$<617$    &$0.36\pm0.05$    &$0.23_{-0.05}^{+0.15}$  &$0.18_{-0.04}^{+0.09}$\\
$\overline{B}_s^0\to \phi\omega$       &$ $    &$0.04\pm0.01$    &$-0.17_{-0.08}^{+0.11}$  &$0.18_{-0.13}^{+0.64}$\\
\hline
\hline
\end{tabular}
\label{bqbr}
\end{table*}

In Table.~\ref{bqbr}, we presented the predictions of the branching fractions, where one finds that our predictions basically agree with the results of PQCD and QCDF.  Similar to $B\to\pi\pi$ modes, $B\to\rho \rho $ decays have been paid more attentions, because they can be used to constrain the CKM angle $\alpha$. Experimentally, the branching fractions of $B\to\rho \rho $ decays exhibit a pattern similar to the corresponding $B\to\pi\pi$ modes. That is, $B^0\to\rho ^+ \rho^-$ and $B^-\to\rho ^- \rho^0$ are nearly equal, while the $B^0\to\rho^0\rho^0$ has a larger branching fraction than naively expected, though far less than $B^0\to\rho ^+ \rho^-$ and $B^-\to\rho^- \rho^0$. In SCET, because $\zeta_J$ is much smaller than $\zeta$, the branching fraction of  $B^0\to\rho^0\rho^0$ is smaller than those of $B^0\to\rho ^+ \rho^-$ and $B^-\to\rho ^- \rho^0$. Obviously, our results are consistent with the experimental data well with a best fitted value of $\zeta_{(J)}$. What needs to be stressed is that the center value of branching fraction of $B^0\to\rho^0\rho^0$ is exactly equal to experimental data. Compared to QCDF and PQCD, the predicted branching fraction of $B^0\to\rho^0\rho^0$ agree with that of QCDF, in which a larger colour-suppressed tree amplitude from the hard spectator scattering are gotten with a large uncertainty. In contrast, in PQCD, the prediction is relatively small because of the large cancellations in the hard-scattering emission diagrams and the annihilation ones. However, for another color-suppressed decay $B^0 \to \rho^0 \omega$, the reverse applies. Our result is in agreement with that of PQCD, and much larger than that of QCDF, because in QCDF the color-suppressed diagrams are almost cancelled by each other, while the extra contributions are from the annihilations in PQCD, as well as the charming penguin in SCET. We also note that the branching fractions of the color-suppressed decays $B_s\to \overline{K}^{*0}\rho(\omega)$ in PQCD are about 3 times smaller than those of QCDF. The decay $B_s\to \overline{K}^{*0}\omega$ in SCET is consistent with PQCD, but more close to QCDF for $B_s\to \overline{K}^{*0}\rho$. For the penguin-dominated decay $B^-\to K^{*0}K^{*-}$, the result of SCET is consistent with results of QCDF and PQCD, but half of the experimental value of BaBar with a relative large uncertainty. We hope the precision of this decay can be improved within the future experiments. For another penguin-dominated decay $\overline{B}_s^0\to K^{*+}K^{*-}$, the predictions from QCDF, PQCD and SCET are consilient with each other, and also agree with the data very well. In term of the flavor SU(3) symmetry,  the branching fractions of $\overline{B}^0\to \rho^+K^{*-}$ and $\overline{B}_s^0\to K^{*+}K^{*-}$ are almost same in this work. We also note that our prediction of the $\overline{B}_s^0\to K^{*+}K^{*-}$ is a bit larger than that of PQCD, and we hope future experimental data could test these approaches. Since we have adopted the flavor SU(3) symmetry in the calculation,  the relations $Br(B^-\to K^{*-}K^{*0})=Br(\overline{B}^0\to \overline{K}^{*0}K^{*0})$, $Br(\overline{B}_s^0\to K^{*+}K^{*-})=Br(\overline{B}^0\to \rho^+K^{*-})$ are obtained. The negligible differences are form the difference of lifetimes of the initial $B$ mesons.  As for decays induced by the QCD flavor-singlet penguin operators, such as $B^0 \to \rho^0 (\omega)\phi$ and $B^- \to \rho ^-\phi$, our results are about one order smaller than those of PQCD, while in QCDF these decays cannot be calculated reliably. Furthermore, due to the smaller branching fractions, these decays are always regarded as probes for searching for possible new physics beyond SM.

Next, we turn to discuss the polarization fractions given in Table.\ref{bqfl}.  Both the color-allowed tree-dominated decays and the QCD flavor-singlet penguin-dominated decays fully respect the helicity amplitude hierarchy in eq.~(\ref{naive fL}) and exhibit predominantly longitudinal polarizations  $f_L \approx 1$ in SCET, PQCD and QCDF. In contrast, those penguin-dominated decays shall spoil the the helicity amplitude hierarchy in eq.~(\ref{naive fL}), and $f_L$ are around $50\%$. These larger transverse polarization fractions for these penguin-dominated decays have already been discussed in previous paragraph. Specially, for $ B^- \to \rho^- \overline{K}^{*0}$, our result is consistent with the data and the prediction of QCDF,  since the parameters used are extracted from this mode. However, in PQCD, the longitudinal  polarization fraction is as large as $70\%$, because in PQCD language this mode has large longitudinal annihilation contribution from tree operators.  As the experimental error is still large, we wait for consolidated data from the Belle-II experiment.

For the color-suppressed tree-dominated decays, from Table.\ref{bqfl}, it is found that their longitudinal  polarization fractions are about   $60 \%$ in both SCET and  PQCD,  except for $B ^0 \to \rho^0\rho^0$ and $\overline{B}_s^0\to K^{*0} \omega$. Conversely, in QCDF, with the exception of $\overline{B}^0\to\rho^0 \omega$, the longitudinal polarizations are dominant, and their fractions are larger than $ 90 \%$.  For $B ^0 \to \rho^0\rho^0$ decay, $f_L$ is about $87\%$ in SCET, which is nearly compatible with experimental data within error bar. However, in PQCD, $f_L$ is as small as $12\%$ \cite{Zou:2015iwa}, because the longitudinal polarization contributions  from two hard-scattering emission diagrams largely cancel against each other in PQCD picture. As aforementioned, in SCET, at leading power, all transverse polarization contributions arise only from charming penguins, the numerical values of which are much smaller than the annihilations appearing in PQCD, which leads to the difference between two predictions. Hereby, it is important to have a refined measurement of the branching fraction and the longitudinal polarization fraction for $B ^0 \to \rho^0\rho^0$ to draw a definitive conclusion.  As for other decays, such as $\overline{B}^0 \to \omega \omega$, the longitudinal polarization fraction is $63\%$, which indicates that the transverse charming penguin gives a significant contribution in SCET.  For the decays $\overline{B}_s^0\to K^{*0}\rho^0$ and  $\overline{B}_s^0\to K^{*0} \omega$, the charming penguins give constructive and destructive contributions to longitudinal polarizations respectively, as seen in Table.\ref{Tid}.  So, it is understandable for us that $f_L(\overline{B}_s^0\to K^{*0}\rho^0)$ is a bit larger than $f_L(\overline{B}_s^0\to K^{*0}\omega)$.

\begin{table*} [!tb]
\centering
\caption{Fractions of the longitudinal polarization $f_L$  of  $ B_{(s)} \to V V$ decays.}
\begin{tabular}{l l l l l }
\hline
\hline
 Channel  & Exp. & SCET & PQCD & QCDF    \\
\hline
$B^-\to\rho^-\rho^0$       &$0.95\pm0.016$    &$1.00$  &$0.98_{-0.01}^{+0.01}$ &$0.96_{-0.02}^{+0.02}$\\
$B^-\to\rho^-\omega$       &$0.90\pm 0.06$    &$0.97\pm0.01$  &$0.97_{-0.01}^{+0.01}$ &$0.96_{-0.03}^{+0.02}$ \\
$B^-\to K^{*0}K^{*-}$  &$0.75_{-0.26}^{+0.16}$    &$0.50\pm0.16$  &$0.74_{-0.05}^{+0.04}$ &$0.45_{-0.38}^{+0.55}$ \\
$B^-\to\rho^-\phi$         &$ $              &$1.00$   &$0.95_{-0.02}^{+0.01}$          &$ $    \\
$B^-\to\rho^0K^{*-}$        &$0.78\pm 0.12$   &$0.42\pm0.14$  &$0.75_{-0.05}^{+0.04}$ &$0.67_{-0.48}^{+0.31}$ \\
$B^-\to\rho^-\overline{K}^{*0}$  &$0.48\pm 0.08$   &$0.45\pm0.18$  &$0.70_{-0.05}^{+0.05}$ &$0.48_{-0.40}^{+0.52}$ \\
$B^-\to K^{*-}\omega$       &$0.41\pm0.19$    &$0.53\pm0.14$  &$0.64_{-0.07}^{+0.07}$ &$0.67_{-0.39}^{+0.32}$ \\
\hline
$\overline{B}^0\to \rho^{+}\rho^{-}$  &$0.988\pm0.026$  &$0.991\pm0.003$  &$0.95_{-0.01}^{+0.01}$ &$0.92_{-0.03}^{+0.01}$ \\
$\overline{B}^0\to \rho^{0}\rho^{0}$  &$0.60_{-0.25}^{+0.21}$  &$0.87\pm0.05$  &$0.12_{-0.02}^{+0.16}$ &$0.92_{-0.37}^{+0.07}$ \\
$\overline{B}^0\to \rho^{0}\omega$     &$ $        &$0.58\pm0.14$  &$0.67_{-0.09}^{+0.08}$ &$0.52_{-0.44}^{+0.51}$ \\
$\overline{B}^0\to \omega\omega$       &$ $   &$0.64\pm0.15$  &$0.66_{-0.11}^{+0.10}$ &$0.94_{-0.20}^{+0.04}$ \\
$\overline{B}^0\to K^{*0}\overline{K}^{*0}$  &$0.80_{-0.13}^{+0.12}$   &$0.50\pm0.16$  &$0.58_{-0.08}^{+0.08}$ &$0.52_{-0.49}^{+0.48}$ \\
$\overline{B}^0\to \rho^{0}\phi$         &$ $     &$1.00$  &$0.95_{-0.01}^{+0.01}$ &$ $             \\
$\overline{B}^0\to \omega\phi$           &$ $      &$1.00$  &$0.94_{-0.03}^{+0.02}$ &$ $             \\
$\overline{B}^0\to \rho^{0}\overline{K}^{*0}$  &$0.40\pm0.14$  &$0.61\pm0.13$  &$0.65_{-0.05}^{+0.04}$ &$0.39_{-0.31}^{+0.60}$ \\
$\overline{B}^0\to \rho^{+}K^{*-}$        &$0.38\pm0.13$  &$0.55\pm0.14$  &$0.68_{-0.05}^{+0.05}$ &$0.53_{-0.32}^{+0.45}$ \\
$\overline{B}^0\to \overline{K}^{*0}\omega$     &$0.69\pm0.13$  &$0.40\pm0.20$  &$0.65_{-0.05}^{+0.05}$ &$0.58_{-0.17}^{+0.44}$ \\
\hline
$\overline{B}_s^0\to K^{*+}\rho^-$  &   &$0.991\pm0.003$  &$0.95_{-0.01}^{+0.01}$  &$0.92_{-0.04}^{+0.01}$\  \\
$\overline{B}_s^0\to K^{*0}\rho^0$  &   &$0.87\pm0.05$    &$0.57_{-0.08}^{+0.09}$  &$0.90_{-0.24}^{+0.05}$\\
$\overline{B}_s^0\to K^{*0}\omega$  &   &$0.64\pm0.15$    &$0.50_{-0.17}^{+0.13}$  &$0.90_{-0.23}^{+0.04}$\\
$\overline{B}_s^0\to K^{*-}K^{*+}$  &      &$0.55\pm0.14$    &$0.42_{-0.11}^{+0.14}$  &$0.52_{-0.22}^{+0.20}$\\
$\overline{B}_s^0\to \phi\rho^0$    &      &$1.00$    &$0.86_{-0.01}^{+0.01}$  &$0.88_{-0.18}^{+0.02}$\\
$\overline{B}_s^0\to \phi\omega$    &      &$1.00$    &$0.69_{-0.13}^{+0.11}$  &$0.95_{-0.42}^{+0.01}$\\
\hline
\hline
\end{tabular}
\label{bqfl}
\end{table*}

In Table.~\ref{bqcp}, we list the predicted direct {\it CP} asymmetries in SCET. For comparison, the experimental data, as well as predictions of PQCD and QCDF, are also presented. Compared with measurements with poor precision,  our results accommodate the experimental data well, except the decay $\overline{B}^0\to \overline{K}^{*0}\omega$.  As is known that  the direct {\it CP} asymmetry depends on both the strong phase and the weak CKM phase. In SCET, only the long-distance charming penguin  $A_{c \overline{c}}$  is able to afford the large strong phase at leading power and leading order in $\alpha_s$ expansion.  Differently,  the  strong phases in QCDF and PQCD are from the hard-scatter and annihilation diagrams. So, in these three approaches, the origins of strong phase are different, which leads to different predictions. From Table.~\ref{bqcp}, it is found that  for the flavor-singlet penguin-dominated decays, such as $B^-\to\rho^-\rho^0$, $B^-\to\rho^-\phi$, $\overline{B}^0\to\rho^0\phi$, $\overline{B}^0\to \omega\phi$,  $\overline{B}_s^0\to\phi\rho^0$ and $\overline{B}_s^0\to\phi\omega$,  the direct {\it CP} asymmetries are zero due to the absence of the charming penguin contributions.  However, the  $\overline{B}_s^0\to\phi\rho^0$ and $\overline{B}_s^0\to\phi\omega$ have larger asymmetries in QCDF and PQCD, because they suffer from large contribution from the hard-scattering diagrams. On the other side, because different methods were used for calculating the hard-scattering and the annihilation diagrams, the signs of the direct {\it CP} asymmetries are different. If experimental data are available, these two decay could be used to differentiate three approaches. We also note  that for the pure penguin-dominated $b\to d$ processes, such as $B^- \to K^{*0}K^{*-}$ and $B^- \to \rho^- K^{*0}$, the direct {\it CP} asymmetries are a little smaller, relative to those of decays induced by $b \to s$ penguin, such as $B^- \to \rho^0 K^{*-}$, $B^- \to K^{*-}\omega$, $B^0 \to \rho^+ K^{*-}$ and $B_s\to K^{*-} K^{*+}$, due to $|V_{td}|<|V_{ts}|$. Moreover,  our predictions of the modes $\overline{B}^0\to \rho^0\rho^0/\omega\omega$ and $\overline{B}^0\to \rho^0\omega$ are consistent with that of  QCDF, but  differ from the results of  PQCD.

\begin{table*} [tbh!]
\centering
\caption{Direct { \it CP} asymmetries ($\%$) of  $ B_{(s)} \to V V$ decays.}
\begin{tabular}{l l l l l}
\hline
\hline
 Channel  & Exp. & SCET &PQCD &QCDF   \\
\hline
$B^-\to\rho^-\rho^0$       &$-5\pm 5$    &$0$  &$0.05_{-0.03}^{+0.06}$  &$0.06$\\
$B^-\to\rho^-\omega$       &$-20\pm 9$    &$-13.6\pm16.1$  &$-11.2_{-3.2}^{+3.1}$  &$-8_{-4}^{+3}$\\
$B^-\to K^{*0}K^{*-}$      &$ $    &$9.5\pm10.6$  &$23.0_{-4.9}^{+4.7}$  &$16_{-34}^{+17}$\\
$B^-\to\rho^-\phi$        &$ $    &$0$  &$0.0$  &$ $\\
$B^-\to\rho^0K^{*-}$        &$31\pm 13$    &$29.3\pm31.0$  &$22.7_{-3.2}^{+2.9}$  &$43_{-28}^{+13}$\\
$B^-\to\rho^-\overline{K}^{*0}$    &$-1\pm 16$    &$-0.56\pm0.61$  &$-1.0_{-0.4}^{+0.3}$  &$-0.3_{-0}^{+2}$\\
$B^-\to K^{*-}\omega$       &$29\pm35$    &$24.3\pm27.1$   &$9.1_{-4.8}^{+3.5}$  &$56_{-43}^{+5}$\\
\hline
$\overline{B}^0\to \rho^{+}\rho^{-}$   &             &$-7.68\pm9.19$  &$0.83_{-0.74}^{+0.83}$  &$-4_{-3}^{+3}$\\
$\overline{B}^0\to \rho^{0}\rho^{0}$  &$ $  &$19.5\pm23.5$  &$70.7_{-9.6}^{+4.8}$  &$30_{-31}^{+22}$\\
$\overline{B}^0\to \rho^{0}\omega$  &$ $  &$8.6\pm10.1$  &$59.4_{-11.8}^{+13.4}$  &$3_{-76}^{+51}$\\
$\overline{B}^0\to \omega\omega$  &$ $  &$-36.8\pm40.1$  &$-73.7_{-8.5}^{+7.9}$  &$-30_{-23}^{+22}$\\
$\overline{B}^0\to K^{*0}\overline{K}^{*0}$  &  &$9.5\pm10.6$  &$0.0$  &$-14_{-2}^{+6}$\\
$\overline{B}^0\to \rho^{0}\phi$  &$ $  &$0$  &$0.0$  &$ $\\
$\overline{B}^0\to \omega\phi$  &$ $  &$0$    &$0.0$  &$ $\\
$\overline{B}^0\to \rho^{0}\overline{K}^{*0}$  &$-6\pm9$  &$-3.30\pm3.91$  &$-8.9_{-3.0}^{+3.0}$  &$-15_{-16}^{+16}$\\
$\overline{B}^0\to \rho^{+}K^{*-}$      &$21\pm15$  &$20.6\pm23.3$  &$24.5_{-3.8}^{+3.1}$  &$32_{-14}^{+2}$\\
$\overline{B}^0\to \overline{K}^{*0}\omega$  &$45\pm25$  &$3.66\pm4.05$  &$5.6_{-1.6}^{+1.5}$  &$23_{-19}^{+10}$\\
\hline
$\overline{B}_s^0\to K^{*+}\rho^-$ &     &$-7.68\pm9.19$    &$-9.1_{-1.9}^{+1.7}$  &$-11_{-1}^{+4}$\\
$\overline{B}_s^0\to K^{*0}\rho^0$ &    &$19.5\pm23.5$       &$62.7_{-18.8}^{+14.4}$  &$46_{-30}^{+18}$\\
$\overline{B}_s^0\to K^{*0}\omega$ &    &$-36.8\pm40.1$    &$-78.1_{-11.3}^{+15.7}$  &$-50_{-16}^{+29}$\\
$\overline{B}_s^0\to K^{*-}K^{*+}$ &       &$20.6\pm23.3$    &$8.8_{-9.4}^{+2.5}$  &$21_{-4}^{+2}$\\
$\overline{B}_s^0\to \phi\rho^0$   &     &$0$    &$-4.3_{-1.2}^{+1.5}$  &$83_{-36}^{+10}$\\
$\overline{B}_s^0\to \phi\omega$   &     &$0$    &$28.0_{-6.4}^{+3.7}$  &$-8_{-15}^{+20}$\\
\hline
\hline
\end{tabular}
\label{bqcp}
\end{table*}

The other predicted seven observables, the perpendicular polarization fraction ($f_\perp$), the relative phases ($\phi_\parallel$, $\phi_\perp$, $\Delta \phi_\parallel$, and $\Delta \phi_\perp$), and {\it CP} asymmetry parameters ($A_{CP}^0$ and $A_{CP}^{\perp}$) in SCET are listed in Table~\ref{bqother}. As analyzed before, the observables with subscript ``$\parallel$ '' are approximately equal to ones with  ``$\perp$''.

\begin{sidewaystable}[!htb]
\centering
\caption{Transverse polarization fractions $f_\perp(\%)$, relative phases $\phi_\parallel$(rad), $\phi_\perp$(rad), $\Delta \phi_\parallel$($10^{-2}$rad), $\Delta \phi_\perp$($10^{-2}$rad) and the {\it CP} asymmetry parameters $A_{CP}^0(\%)$ and $A_{CP}^\perp(\%)$ of $B_{(s)}\to V V$ decays.}
\small\begin{tabular}{l l l l l l l l}
\hline
\hline
 Channel  & $f_\perp$ & $\phi_{\parallel}$ &$\phi_{\perp}$ &$A_{CP}^0$ &$A_{CP}^{\perp}$ &$\Delta\phi_{\parallel}$   &$\Delta\phi_{\perp}$\\
\hline
$B^-\to\rho^-\omega$       &$1.27\pm0.56 $ &$-1.18\pm0.35$   &$-1.05\pm0.35$  &$-0.39\pm0.50$  &$13.6\pm16.1$  &$137\pm6$  &$137\pm6$\\
$B^-\to K^{*0}K^{*-}$      &$22.9\pm10.0$    &$2.41\pm0.63$  &$2.54\pm0.63$  &$9.7\pm11.5$  &$-9.5\pm10.6$  &$20.6\pm8.8$ &$20.6\pm8.8$\\
$B^-\to\rho^0K^{*-}$        &$26.2\pm 9.9$    &$2.33\pm0.61$  &$-0.68\pm0.61$  &$40.4\pm51.3$  &$-29.3\pm31.0$  &$76.2\pm32.9$ &$-238\pm33$\\
$B^-\to\rho^-\overline{K}^{*0}$  &$24.9\pm 11.1$    &$2.47\pm0.67$ &$2.60\pm0.67$  &$-0.68\pm0.77$  &$0.56\pm0.61$   &$-1.17\pm0.60$ &$-1.17\pm0.60$\\
$B^-\to K^{*-}\omega$       &$21.3\pm8.7$    &$2.31\pm0.55$ &$2.44\pm0.55$  &$21.5\pm28.3$  &$-24.3\pm27.1$   &$59.7\pm19.6$ &$59.7\pm19.6$\\
\hline
$\overline{B}^0\to \rho^{+}\rho^{-}$  &$0.40\pm0.18$  &$-1.17\pm0.35$ &$-1.04\pm0.35$ &$-0.07\pm0.09$  &$7.68\pm9.19$  &$132\pm3$ &$132\pm3$\\
$\overline{B}^0\to \rho^{0}\rho^{0}$  &$5.81\pm2.84$  &$2.09\pm0.38$  &$2.22\pm0.38$  &$2.87\pm4.00$  &$-19.5\pm23.5$  &$77.7\pm11.0$ &$77.7\pm11.0$\\
$\overline{B}^0\to \rho^{0}\omega$  &$19.0\pm8.4$  &$2.34\pm0.56$ &$2.47\pm0.56$ &$6.25\pm7.83$  &$-8.6\pm10.1$ &$19.5\pm7.6$ &$19.5\pm7.6$\\
$\overline{B}^0\to \omega\omega$  &$16.5\pm8.4$  &$-1.09\pm0.37$ &$-0.96\pm0.37$ &$-21.0\pm32.4$  &$36.8\pm40.1$   &$180\pm28$ &$180\pm28$\\
$\overline{B}^0\to K^{*0}\overline{K}^{*0}$  &$22.9\pm10.0$  &$2.41\pm0.63$ &$2.54\pm0.63$ &$9.7\pm11.5$  &$-9.5\pm10.6$   &$20.6\pm8.8$ &$20.6\pm8.8$\\
$\overline{B}^0\to \rho^{0}\overline{K}^{*0}$  &$17.6\pm7.9$  &$2.32\pm0.54$ &$2.45\pm0.54$ &$-2.10\pm2.67$  &$3.30\pm3.91$   &$-7.55\pm2.69$ &$-7.55\pm2.69$\\
$\overline{B}^0\to \rho^{+}K^{*-}$      &$20.3\pm8.6$  &$2.32\pm0.55$ &$2.45\pm0.55$  &$16.8\pm21.7$  &$-20.6\pm23.3$   &$49.1\pm15.9$ &$49.1\pm15.9$\\
$\overline{B}^0\to \overline{K}^{*0}\omega$  &$27.0\pm12.0$  &$2.52\pm0.72$ &$2.65\pm0.72$  &$5.42\pm5.91$  &$-3.66\pm4.05$  &$7.58\pm4.98$ &$7.58\pm4.98$\\
\hline
$\overline{B}_s^0\to K^{*+}\rho^-$      &$0.40\pm0.18$    &$-1.17\pm0.35$ &$-1.01\pm0.35$  &$-0.07\pm0.09$  &$7.68\pm9.19$  &$132\pm3$ &$132\pm3$\\
$\overline{B}_s^0\to K^{*0}\rho^0$     &$5.81\pm2.84$    &$2.09\pm0.38$ &$2.22\pm0.38$ &$2.87\pm4.00$  &$-19.5\pm23.5$  &$77.7\pm11.0$ &$77.7\pm11.0$\\
$\overline{B}_s^0\to K^{*0}\omega$     &$16.5\pm8.4$    &$-1.09\pm0.37$ &$-0.96\pm0.37$ &$-21.0\pm32.4$  &$36.8\pm40.1$  &$180\pm28$ &$180\pm28$\\
$\overline{B}_s^0\to K^{*-}K^{*+}$        &$20.3\pm8.6$    &$2.32\pm0.55$ &$2.45\pm0.55$ &$16.8\pm21.7$  &$-20.6\pm23.3$  &$49.1\pm15.9$ &$49.1\pm15.9$\\
\hline
\hline
\end{tabular}
\label{bqother}
\end{sidewaystable}

Under QCDF and PQCD framework, the U-spin symmetry works well \cite{Cheng:2008gxa, Zou:2015iwa}.  Now, we also examine the U-spin symmetry in SCET according to the following relations:
\begin{eqnarray}
&&A_{CP}(\overline{B}_s^0\to K^{*+}\rho^-)=-A_{CP}(\overline{B}^0\to K^{*-}\rho^+)\frac{Br(\overline{B}^0\to K^{*-}\rho^+)}{Br(\overline{B}_s^0\to K^{*+}\rho^-)}\frac{\tau(B_s)}{\tau(B)},\nonumber\\
&&A_{CP}(\overline{B}_s^0\to K^{*0}\rho^0)=-A_{CP}(\overline{B}^0\to \overline{K}^{*0}\rho^0)\frac{Br(\overline{B}^0\to \overline{K}^{*0}\rho^0)}{Br(\overline{B}_s^0\to K^{*0}\rho^0)}\frac{\tau(B_s)}{\tau(B)},\nonumber\\
&&A_{CP}(\overline{B}_s^0\to K^{*+}K^{*-})=-A_{CP}(\overline{B}^0\to \rho^{-}\rho^+)\frac{Br(\overline{B}^0\to \rho^-\rho^+)}{Br(\overline{B}_s^0\to K^{*+}K^{*-})}\frac{\tau(B_s)}{\tau(B)},\nonumber\\
&&A_{CP}(\overline{B}_s^0\to K^{*0}\overline{K}^{*0})=-A_{CP}(\overline{B}^0\to K^{*0}\overline{K}^{*0})\frac{Br(\overline{B}^0\to K^{*0}\overline{K}^{*0})}{Br(\overline{B}_s^0\to K^{*0}\overline{K}^{*0})}\frac{\tau(B_s)}{\tau(B)}.
\end{eqnarray}
In Table.~\ref{buspin}, we list the numerical results of {\it CP} asymmetries based on U-spin symmetry, together with the values of the direct calculation in SCET.  From the table, one can find that  the U-spin symmetry also works well in SCET framework with certain uncertainties.

\begin{table*}[!tbh]
\centering
\caption{The direct CP asymmetry in $\overline{B}_s^0\to VV$ decays via U-spin symmetry and the direct SCET calculation}\label{buspin}
{\small
\begin{tabular}{l l l l l l l}
\hline
\hline
 Modes  & Br($10^{-6}$) & $A_{CP}(\%)$ &   Modes &Br($10^{-6}$)  & $A_{CP}(\%)$(U-spin) &$A_{CP}^{\%}$(SCET) \\
\hline
$\overline{B}^0\to\rho^+K^{*-}$  &$10.6 $ &$20.6\pm23.3$   &$\overline{B}_s^0\to K^{*+}\rho^-$  &$28.8$       &$-7.58$    &$-7.68\pm9.19$\\
$\overline{B}^0\to\rho^0\overline{K}^{*0}$   &$5.87 $ &$-3.3\pm3.9$   &$\overline{B}_s^0\to K^{*0}\rho^0$  &$1.04$  &$18.6$   &$19.5\pm23.5$\\
$\overline{B}^0\to\rho^+\rho^{-}$  &$27.7 $ &$-7.68\pm9.19$   &$\overline{B}_s^0\to K^{*-}K^{*+}$  &$11.0$  &$19.3$    &$20.6\pm23.3$\\
$\overline{B}^0\to K^{*0}\overline{K}^{*0}$  &$0.48 $ &$9.5\pm10.6$   &$\overline{B}_s^0\to K^{*0}\overline{K}^{*0}$  &$8.60$  &$-0.53$    &$-0.56\pm0.61$\\
\hline
\hline
\end{tabular}
}
\end{table*}
\section{Conclusion}\label{Sec-4}
In this work, we have studied the charmless $B\to VV$ decays at leading power in $1/m_b$ and leading order in $\alpha_s$ in   soft-collinear effective theory. In the flavor SU(3) symmetry limit,   there are   8 unknown nonperturbative parameters, which are obtained by  fitting  35 observables  in the  minimum $\chi^2$ approach.  The obtained results have indicated that the hard-scattering form factor is much smaller than the soft form factor $\zeta_J^V\ll\zeta^V$. This character differs significantly  from the relation $\zeta_J^P\simeq\zeta^P$ for the $B$-to-pseudo-scalar meson derived in a global analysis of   $B\to PP(V)$ decays. Using the best fitted parameters,  we have calculated all observables for the 28 decay modes, including   branching fractions, polarization fractions, relative phases, and  direct {\it CP} asymmetries. Most of our results are   compatible with the present experimental data when available, while the  other predictions  can  be tested in the ongoing LHCb experiment and the forthcoming Belle-II experiment. Furthermore, the nonperturbative and  universal charming penguins $A_{cc(L,\parallel,\perp)}$  spoil the naive helicity relation, and they can be used to explain the large transverse polarization fractions observed in the experiments. In addition, the large strong phases in the charming-penguins are responsible for direct $CP$ asymmetries for some decays. Because the origins of strong phases are different in QCDF, PQCD and SCET, the high precision study of direct $CP$ asymmetries in the future   can help us to discriminate these three approaches.

\section*{Acknowledgments}
We are very grateful to Wei Wang for helpful discussion and reading the manuscript. The work is partly supported by National Natural Science Foundation of China (11575151, 11375208, 11521505, 11621131001 and 11235005).   Y.Li is also supported by the Natural Science Foundation of Shandong Province (Grant No.ZR2016JL001).

\end{document}